\documentclass[12pt]{article}
\usepackage{graphicx}
\begin{document}
Justification of Sexual Reproduction by Modified Penna Model of Ageing
\footnote{Dedicated to E. M\"uller-Hartmann on the occasion of his 60th 
birthday}
\bigskip

J.S. S\'a Martins$^1$ and D. Stauffer$^{2,3}$
\bigskip

$^1$  Colorado Center for Chaos and Complexity/CIRES, 

University of Colorado, Boulder CO 80309, USA

$^2$ Instituto de F\'{\i}sica, Universidade Federal Fluminense, 
  
Av. Litor\^anea s/n, Boa Viagem, Niter\'oi 24210-340, RJ, Brazil 

\medskip
$^3$visiting from 
Inst. for Theoretical Physics, Cologne University, D-50923 K\"oln, Euroland

\medskip
e-mail: jorge@cires.colorado.edu, stauffer@thp.uni-koeln.de

\bigskip
Abstract: We generalize the standard Penna bit-string model of biological ageing
by assuming, that each deleterious mutation diminishes the survival probability
in every time interval by a small percentage. This effect is added to the 
usual lethal but age-dependent effect of the same mutation. We then
find strong advantages or disadvantages of sexual reproduction (with males and
females) compared to asexual cloning, depending on parameters.


\bigskip
Presently, animals on earth proliferate mainly in two different ways: Sexually
with separate males and females (now abbreviates by SX), and asexually with
only one gender (abbreviated by AS). Less widespread are intermediate or
mixed forms, like hermaphroditism. SX and AS coexist stably since hundreds of 
million years, with mammals only proliferating sexually and bacteria mostly
cloning asexually. 

In the Redfield model \cite{redfield,stauffer},
computer simulations have given clear advantages of one or the other way,
depending on the parameters (even
if we include that males fail to get pregnant and thus reduce the average 
birthrate per animal by a factor of two). A genomic bitstring model without
ageing
\cite{turkey} also justified sexual reproduction; in one case the asexual
way of life even died out. No such clear advantage was found until now in the 
more realistic Penna ageing model \cite{penna}; there an advantage of
AS over SX was seen, which turned into a slight disadvantage for AS if compared
to hermaphroditism and meiotic parthenogenesis
\cite{bernardes,stamm,anais}. Such slight (dis)\-advantages can easily be 
reversed by 
minor effects not included in the simulated model, like the effort to find
a sexual partner. Thus now we present a modification of the Penna model which
gives drastic advantages for SX compared with AS, or the opposite result if
a single parameter is changed.

In agreement with many earlier models, including \cite{redfield}, we assume that
each mutation reduces the survival probability by a small fraction
$\epsilon$. Thus at each iteration or ``year"", each animal survives with
probability exp($-m\epsilon$) if it has $m$ mutations. All mutations are assumed
hereditary and detrimental, as is customary in ageing theories \cite{wachter}.
In addition, the same mutations have the lethal age-dependent effects of the 
standard Penna model \cite{penna,book}, and a Verhulst factor acting on all
ages takes into account the limits of food and space. Thus a mutation,
realized e.g. by a bit set to one in position $x$ of the bit-string, causes
a life-threatening disease at ages $x$ and later, as well as a small additional
mortality $\epsilon$ per year for all ages. This possibility of one gene having 
different effects is known as pleiotropy \cite{almeida,sousa}; in contrast
to other ageing theories we do not have to assume antagonistic pleiotropy where
mutations have both positive and negative effects at different ages. 

\begin{figure}[hbt]
\begin{center}
\includegraphics[angle=-90,scale=0.5]{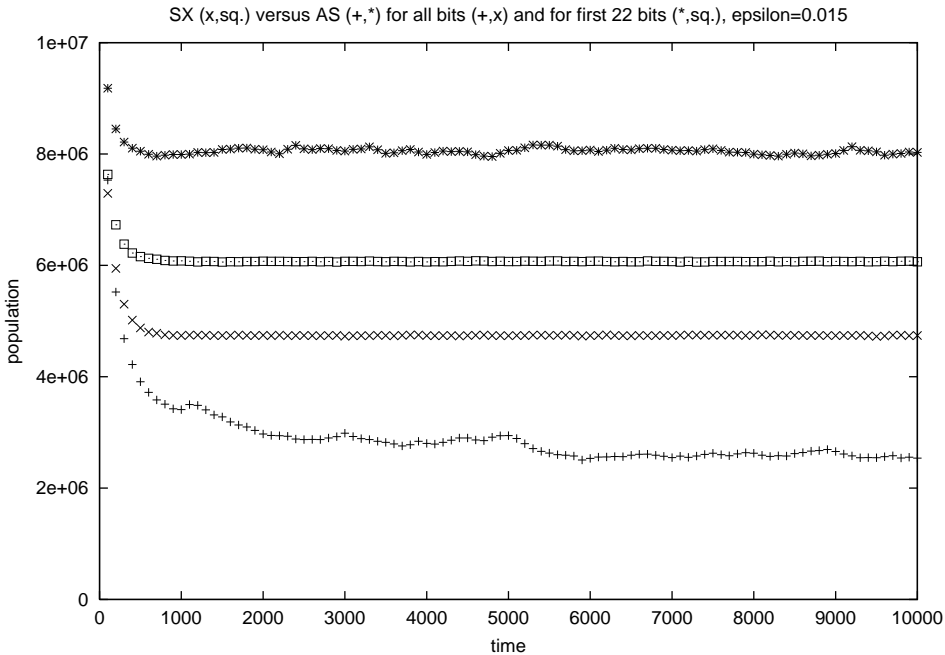}
\end{center}
\caption{
Advantage of sexual (x) versus asexual (+) reproduction in pleiotropic
Penna model. The effect is inverted (squares for SX, stars for AS) if the ten 
bits corresponding to the oldest ages are ignored in the additional mortality.
}
\end{figure}    

The Fortran programs follow the published ones \cite{book} and are available 
from the authors; the birth rate for the females in SX and for all animals in 
AS was four, while $\epsilon=0.015$. For SX, mutations count only if they are 
inherited from both father and mother, or if they happen on a dominant locus and
are inherited from one of the two parents; for AS, all mutations count.
Our Fig.1 shows that the sexual population is twice as high as the
asexual one, in separate simulations. (In a simultaneous simulation of
two populations, the one which in separate simulations with the same 
parameters gives the higher stationary population \cite{stamm} drives
the other to extinction, as we tested in related models.) Unfortunately,
the SX mortalities deviate for these parameters stronger than usual 
\cite{book} from the exponential Gompertz increase with age.

The situation changes drastically, in favour of AS, if the mutations in the
last 10 of 32 bits (corresponding to the last ten age intervals) are ignored
in the mortality exp($-m\epsilon$). These curves are also shown in our figure.
Thus, depending on biological details, either SX or AS can be the clearly 
preferred choice, in agreement with reality.

\begin{figure}[hbt]
\begin{center}
\includegraphics[angle=-90,scale=0.5]{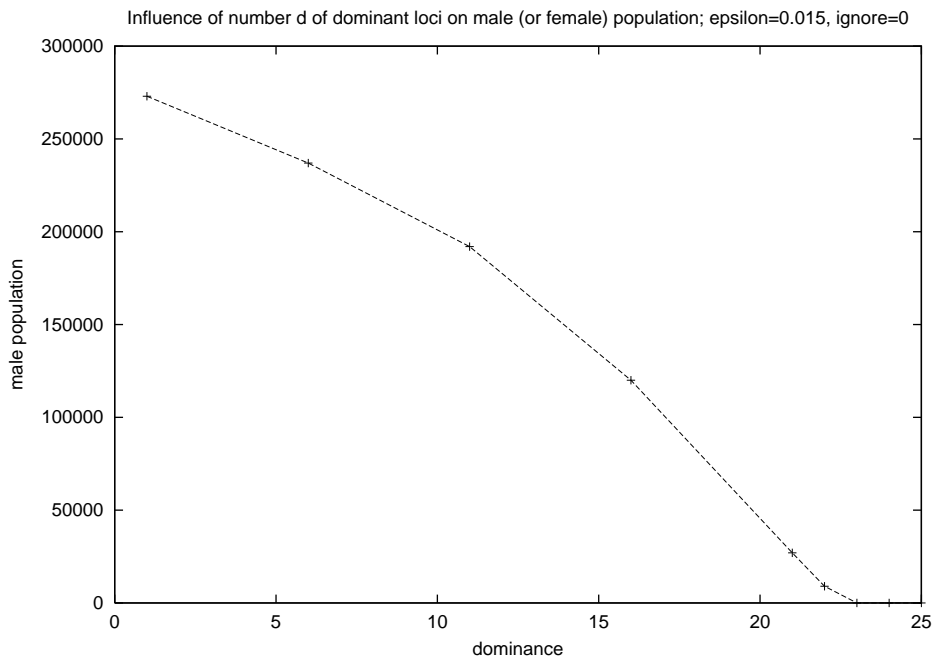}
\end{center}
\caption{
Variation of male (or female) population as a function of the number $d$
of dominant loci for SX. Fig.1 uses $d=6$ and ten times better statistics. For
$d \ge 23$ the whole population died out.
}
\end{figure}    

With no mutations ignored, Fig.2 shows the drastic influence of the number $d$
of dominant loci on the SX population; in the traditional Penna SX model this
influence is much weaker. 

In summary, finally also for the Penna ageing model conditions were found where
sexual reproduction is clearly preferred over asexual cloning. This may help
to explain the origin of sex $10^9$ years ago, without external effects
(questioned in \cite{stamm})
like parasites \cite{lively,martins} or catastrophes \cite{dasgupta,marmoss}.

We thank S. Moss de Oliveira for a critical reading of the manuscript, and her
and P.M.C. de Oliveira for hospitality extended to DS.


\end{document}